\documentclass[11pt]{article}
\usepackage{epsfig,psfig,float,amssymb,latexsym}
\newcommand{\NP}[1]{ Nucl.\ Phys.\ {\bf #1}}

\newcommand{\PL}[1]{ Phys.\ Lett.\ {\bf #1}}

\newcommand{\AN}[1]{Ann. Phys. NY {\bf #1}}

\newcommand{\PR}[1]{Phys.\ Rev.\ {\bf #1}}
\newcommand{\PRL}[1]{ Phys.\ Rev.\ Lett.\ {\bf #1}}

\newcommand{\bi}{\bibitem}
\newcommand{\La}{{\cal L}}

\newcommand{\Opd}{\mathcal{O}(p^2)}
\newcommand{\Opc}{\mathcal{O}(p^4)}
\newcommand{\Ima}{{\rm Im}\,}

\newcommand{\be}{\begin{equation}}
\newcommand{\ee}{\end{equation}}
\newcommand{\ba}{\begin{eqnarray}}
\newcommand{\ea}{\end{eqnarray}}

\begin{document}

\begin{center}
{\large{\bf The Case of a WW Dynamical Scalar Resonance within a Chiral 
Effective Description of the Strongly Interacting Higgs Sector. }}
\end{center}
\vspace{.5cm}

\begin{center}
{\large{J. A. Oller}}\footnote{Present address: {\small{\it Forschungzentrum 
J\"ulich, Institut f\"ur Kernphysik (Theorie), 
D-52425 J\"ulich, Germany}}}
\end{center}

\begin{center}
{\small{\it Departamento de F\'{\i}sica Te\'orica and IFIC\\ Centro Mixto 
Universidad de Valencia-CSIC\\
46100 Burjassot (Valencia), Spain}}
\end{center}
\vspace{1cm}

\begin{abstract}
{\small{We have studied the strongly interacting $W_L W_L \rightarrow W_L W_L$
I=L=0 partial wave amplitude making use of effective chiral Lagrangians. The
Higgs boson is explicitly included and the N/D method is used to unitarize the
amplitude. We recover the chiral perturbative expansion at next to leading 
order for low energies. The cases $m_H<< 4\pi v$ and $m_H>> 4 \pi v$  are 
considered in detail. It is shown that in the latter situation a state appears 
with a mass $\lesssim 1$ TeV. This state is dynamically generated through the 
strong interactions 
between the $W_L$ and is not responsible for the spontaneous
electroweak symmetry breaking. However, its shape can be very similar 
to that of the $m_H\lesssim 1$ TeV case, which corresponds to a conventional
heavy Higgs boson.}}  
\end{abstract}

\vspace{0.5cm}

Although the $SU(2)_L\otimes U(1)_Y$ theory of electroweak interactions is
extremely successful, its spontaneous breaking to $U(1)_{em}$ is a controversial
aspect. In the so-called standard model (SM) this is accomplished
through a Higgs boson which provides a renormalizable way to generate the needed
$W$ and $Z$ masses. However, an experimental verification of this mechanism is
still lacking. In fact, unveiling the nature of the electroweak symmetry 
breaking
sector is the number one aim at the LHC which could detect the Higgs boson up
to a mass $m_H\lesssim 1$ TeV, with $m_H$ the mass of the Higgs boson.

Making use of an effective theory formalism \cite{AB,L,GLNY} we 
will study in this work the strongly interacting Higgs sector, $m_H\gtrsim 1$
 TeV. Throughout the paper the SM will be considered first. Then, the 
extension of our conclusions to more general scenarios will be also discussed.

The effective theory formalism is suited to study the strong electroweak 
Goldstone boson interactions since for energies well above the mass of the weak
bosons ($m_W\approx 0.1$ TeV) their scattering amplitudes coincide with those of
the longitudinal components of the electroweak gauge bosons ($W_L$). This is
the content of the so-called equivalence theorem \cite{4dGeorgi}. 

We will consider the class of models in which the symmetry breaking sector has a
chiral $SU(2)_L\otimes SU(2)_R$ symmetry that breaks spontaneously to the
diagonal $SU(2)_{L+R}$ subgroup. The latter ``custodial" SU(2) is sufficient
\cite{6dGeorgi} (but has not been proved necessary) to protect 
$\rho=\frac{M_W^2}{M_Z^2 \cos{\theta_W}}\simeq 1$ \cite{6dJapon1}. This symmetry
together with the equivalence theorem guarantees also the universal scattering
theorems for the strongly interacting $W's$ and $Z's$ \cite{Georgi}. The former 
symmetry breaking pattern is the one
found in the SM for $g'=0$, with $g'$ the coupling associated to the
hypercharge. In the case of chiral perturbation theory ($\chi PT$) \cite{GLNY} 
one has the same symmetry breaking situation than before but in this case the 
custodial symmetry is called isospin (I). 

We will consider the $\Opd$ and $\Opc$ chiral Lagrangians derived in
\cite{AB,L}. The equivalence theorem can be reconciled with this low energy
expansion \cite{Pela} as long as we keep just the lowest order in $g$, which is
a rather good approximation. Nevertheless, as we are interested only in the 
scattering of the Goldstone bosons assuming the custodial $SU(2)$ symmetry we 
only need the following terms: 

\ba
\label{La}
\La_2&=&\frac{v^2}{4}<\partial_\mu U \partial^\mu U^\dagger> \\ \nonumber 
\La_4&=&L_1\, <\partial_\mu U \partial^\mu U^\dagger>^2+L_2 \,<\partial_\mu U
\partial^\nu U^\dagger>^2
\ea
where
\be
\label{U}
U=\exp(\frac{i\pi^i\tau^i}{v})
\ee
with $\tau^i$ the Pauli matrices, $\pi^i$ the Goldstone bosons,
$v=(\sqrt{2}\,G_F)^{-1/2}\simeq\frac{1}{4}$ TeV and the symbol $<>$ represents
the trace of the matrix inside it.

The Lagrangian given in eq. (\ref{La}) describes the scattering of massless
pions in $\chi PT$ \cite{GLNY} up to $\Opc$ just by changing $v$ to 
$f_\pi=92.4$ MeV. Keeping in mind the equivalence theorem, we can write the 
$W_L\,W_L$ scattering amplitude up to $\Opc$
from the $\chi PT$ calculation \cite{GLNY} in the chiral limit (massless pions).
The result is:
\ba
\label{Astu}
A(s,t,u)&=&\frac{s}{v^2}+
\frac{15 s^2+7(t-u)^2}{576 \pi^2 v^4}+\frac{2}{v^4}
\left[(4L_1^r(\mu)+L_2^r(\mu))s^2+L_2^r(\mu)(t-u)^2 \right] \\ \nonumber 
&-&\frac{1}{96 \pi^2 v^4}
\left[3s^2\log\frac{-s}{\mu^2}+t(t-u)\log\frac{-t}{\mu^2}+u(u-t)\log
\frac{-u}{\mu^2} \right]
\ea
with $L_i^r(\mu)$ the finite renormalized value of $L_i$ at the scale $\mu$ such
that

\be
\label{li}
L_i=L_i^r(\mu)+\gamma_i\, \lambda
\ee
where
\be
\label{lambda}
\lambda=\frac{\mu^{d-4}}{16 \pi^2} \left[
\frac{1}{d-4}-\frac{1}{2}(\log(4\pi)+\Gamma'(1)+1)\right]
\ee
and 
\ba
\label{gamma}
\gamma_1&=&\frac{1}{12} \\ \nonumber
\gamma_2&=&\frac{1}{6}
\ea

In deriving eq. (\ref{Astu}) from that of \cite{GLNY} we have multiplied by 4
the term with the counterterms $L_i^r(\mu)$ in order to connect with the work
\cite{Herrero} where the counterterms are calculated in the SM for a heavy
Higgs. From this work one has:

\ba
\label{counter}
L_1^r(\mu)&=&\frac{v^2}{8\,m_H^2}-\frac{1}{16 \pi^2} \frac{1}{24}\left(
\frac{76}{3}-\frac{27 \pi}{2\sqrt{3}}-\log\frac{m_H^2}{\mu^2}\right) \\ 
\nonumber
L_2^r(\mu)&=&-\frac{1}{16 \pi^2}\frac{1}{12}\left( \frac{11}{6}-
\log\frac{m_H^2}{\mu^2}\right)
\ea

If the $L_i^r(\mu)$ in eq. (\ref{Astu}) are substituted by those of eq.
(\ref{counter}) one can see that the $\mu$ dependence of $A(s,t,u)$ disappears
and the scale is then fixed by $m_H$. The resulting amplitude would reproduce
the one loop ${\mathcal{O}}(G_F\,m_H^2)$ amplitude in the standard model first 
calculated in refs. \cite{refe} and valid for $4\pi v$, $m_H>>\sqrt{s}>>m_W$.

The term ${\displaystyle{\frac{v^2}{8\,m_H^2}}}$ in $L_1^r(\mu)$ in eq. 
(\ref{counter}) comes
from the exchange at tree level of the Higgs bosons at $\Opc$. In fact, the
contribution of the tree level Higgs to $A(s,t,u)$ to all orders can be easily 
calculated and is given by:
\be
\label{Hex}
\frac{s^2}{v^2(m_H^2-s)}
\ee

The former result up to $\Opc$ coincides with the contribution of the Higgs
boson to $A(s,t,u)$ coming from the term $v^2/(8 \, m_H^2)$ in $L_1^r(\mu)$ of
eq. (\ref{counter}). We will then make a resummation of counterterms up to an
infinite order by replacing the $\Opc$ Higgs exchange contribution by the result
given in eq. (\ref{Hex}) with the full propagator. Substituting also the rest of
eq. (\ref{counter}) in eq. (\ref{Astu}) one finally obtains:

\ba
\label{Astu2}
A(s,t,u)&=&\frac{s}{v^2}+\frac{s^2}{v^2(m_H^2-s)}-\frac{1}{96 \pi^2 v^4}\left[
s^2(50-\frac{27\pi}{\sqrt{3}})+\frac{2}{3}(t-u)^2 \right]\\ \nonumber
&-&\frac{1}{96 \pi^2 v^4}
\left[3s^2\log\frac{-s}{m_H^2}+t(t-u)\log\frac{-t}{m_H^2}+u(u-t)\log
\frac{-u}{m_H^2} \right]
\ea

We will concentrate in the I=0 S-wave partial amplitude, $T(s)$, the one which
corresponds to the Higgs boson. In terms of $A(s,t,u)$ this partial wave is
given by:

\be
\label{pw}
T(s)=\frac{1}{4}\int_{-1}^{1} d\cos{\theta} \left(
3\,A(s,t,u)+A(t,s,u)+A(u,t,s)\right)
\ee

With this definition elastic unitarity reads for $s>0$:

\be
\label{eu}
\Ima{T(s)}=\frac{1}{16\pi} |T(s)|^2
\ee
or equivalently
\be
\label{eu2}
\Ima{1/T(s)}=-\frac{1}{16 \pi}
\ee

Taking into account eqs. (\ref{Astu2}) and (\ref{pw}) one has:

\ba
\label{T}
T(s)&=&\frac{s}{v^2}+\frac{3 s^2}{2 v^2 (m_H^2-s)}+\frac{m_H^4}{s
v^2}\left[\log (1+\frac{s}{m_H^2})-\frac{s}{m_H^2}+\frac{s^2}{2 \,m_H^4}\right] 
\\ \nonumber
&-&\frac{s^2}{1728 \pi^2 v^4}\left[ 1673-297\sqrt{3} \pi+108 \log
\frac{-s}{m_H^2}+42 \log \frac{s}{m_H^2}\right] 
\ea
where we have added and subtracted the $\mathcal{O}(p^0)$ and $\Opd$
contributions of $m_H^4/(s \,v^2) \log(1+s/m_H^2)$, due to the exchange of
the Higgs in the crossed channels. In this way, the fact that any 
exchange of the Higgs boson begins at $\Opc$, as it is clear from eq. 
(\ref{Hex}), is explicitly shown.

In order to study the resonance spectrum of the strongly interacting Higgs
sector we are going to make an N/D \cite{Chew} representation of the former
chiral perturbative result. In a former work \cite{nd}, we used this method
to study the strong
meson-meson scattering including the resonance region, which in this case
appears typically for $\sqrt{s}\gtrsim 1$ GeV.
In the N/D method a partial wave amplitude is 
expressed as the quotient of two functions,

\begin{equation}
\label{n/d}
T(s)=\frac{N(s)}{D(s)}
\end{equation}
with the denominator function $\hbox{D}(s)$, bearing the right hand cut or
unitarity cut and the numerator function $\hbox{N}(s)$, the left hand cut.
 
Taking into account eq. (\ref{eu2}), $N(s)$ and 
$D(s)$ will obey the following equations:

\begin{equation}
\label{eqs1}
\begin{array}{ll}
\Ima  D(s)=\Ima T(s)^{-1}\; N=-\frac{1}{16\pi} \,N(s) &  s>0 \\
\Ima D(s)=0 &  s<0
\end{array}
\end{equation}
\begin{equation}
\label{eqs2}
\begin{array}{ll}
\Ima  N(s)=\Ima T(s)\; D(s)\equiv \Ima T_{Left} \; D(s) & s<0  \\
\Ima N(s)=0 & s>0
\end{array}
\end{equation}

 In ref. \cite{nd} we do not include the left hand cut, although some
 estimations were done. In eq. (\ref{T}) the 
 left hand cut first appears through the term 
 ${\displaystyle \log \frac{s}{m_H^2}}$ which acquires an imaginary part for $s<0$. However, in order
to reproduce $T(s)$ up to the order calculated in eq. (\ref{T}), we will 
consider here the left hand cut in a
perturbative way, such that our $N$ function will satisfy eq. (\ref{eqs2}) up to
 one loop calculated at $\Opc$.

When no left hand cut is included one can always take \cite{nd}:

\ba
\label{nlhc}
N(s)&=&1 \\ \nonumber
D(s)&=&\widetilde{a}+\sum_{i}\frac{R_i}{s-s_i}+g(s)\equiv K^{-1}(s)+g(s)
\ea
where each term of the sum is referred to as a CDD pole \cite{CDD} and $g(s)$ 
is given by

\be
\label{gs}
g(s)=\frac{1}{16 \pi^2}\left( \log \frac{-s}{m_H^2}-a\right)
\ee

In \cite{nd} we prove that the form for $D(s)$ in eq. (\ref{nlhc}) has enough
room to accommodate the exchange in the s-channel of S and P-wave resonances
plus polynomials terms. \footnote{In ref. \cite{nd} the prove was restricted to
the case when the polynomial terms come from the $\La_2$ chiral
Lagrangian. The extension of the proof to include also $\Opc$ local terms can be
done in a straightforward way.} This is exactly the situation we have 
from eq. (\ref{T}), when removing those terms, namely, 
${\displaystyle{\log\frac{s}{m_H^2} 
\hbox{ and } \left[\log (1+\frac{s}{m_H^2})-\frac{s}{m_H^2}+\frac{s^2}{2
\,m_H^4}\right]}}$ which give rise to the left hand cut.

In eq. (\ref{nlhc}) elastic unitarity is fulfilled to all orders in the chiral
expansion. For $s>16 m_W^2$ we have neglected the multi $W_L$ channels, with 
four or more $W_L$. In fact, in eq. (\ref{T}) only 2$W_L$ appear in the loops 
since the inclusion of 4$W_L$ requires at least two loops, which is
$\mathcal{O}(p^6)$.

Expanding $T(s)$ from eq. (\ref{nlhc}) up to one loop at $\Opc$ and comparing 
with eq.(\ref{T}) without those terms responsible for the left hand cut given 
above, one has:

\be
\label{k}
K-K_2^2\,g(s)=\frac{s}{v^2}+\frac{3 s^2}{2\,v^2(m_H^2-s)}-\frac{s^2}{1728 \pi^2
v^4}\left(1673-297 \sqrt{3} \pi\right)-\frac{s^2}{16 \pi^2
v^4}\log\frac{-s}{m_H^2}
\ee
where $K_2$ is the $K$ function at $\Opd$. Hence:

\be
\label{kk}
K(s)=\frac{s}{v^2}+\frac{3s^2}{2 v^2(m_H^2-s)}-\frac{s^2}{1728 \pi^2 v^4}\left(
1673-297 \sqrt{3} \pi+a \right)
\ee

In order to take into account the crossed channel contributions present in eq.
(\ref{T}) we write

\ba
\label{n}
N(s)&=&1+\delta N(s) \\ \nonumber
D(s)&=&K(s)^{-1}+g(s)\left(1+\delta N(s)\right)
\ea

In this way, the resulting $T(s)=N(s)/D(s)$ I=L=0 partial wave will satisfy
unitarity to all orders. Expanding $T(s)$ up to one loop at $\Opc$ and comparing
the result with eq. (\ref{T}), taking into account also eq. (\ref{k}), one has:

\be
\label{deltaN}
\delta N(s)=K^{-1}(s)\left[-\frac{7 s^2}{288 \pi^2
v^4}\log\frac{s}{m_H^2}+\frac{m_H^4}{v^2 s}\left(
\log(1+\frac{s}{m_H^2})-\frac{s}{m_H^2}+\frac{s^2}{m_H^4}\right) \right]
\ee

Hence, our final amplitude $T(s)$ will read

\ba
\label{Tf}
T(s)&=&\frac{1+\delta N(s)}{K^{-1}(s)+g(s)\left(1+\delta N(s)\right)}\\
\nonumber
&=&\frac{1}{\left[K\left(1+\delta N(s)\right)\right]^{-1}+g(s)}
\ea

From eqs. (\ref{kk}) and (\ref{deltaN}) one has:

\ba
\label{kdeltaN}
K(s)\left(1+\delta N(s)\right)&=&\frac{s}{v^2}-\frac{s^2}{1728 
\pi^2 v^4}\left(1673-297 \sqrt{3}\pi+108 \,a+42\,
\log \frac{s}{m_H^2} \right)\\
\nonumber
&+&\frac{3 s^2}{2 v^2
(m_H^2-s)}+\frac{m^4}{v^2
s}\left(\log(1+\frac{s}{m_H^2})-\frac{s}{m_H^2}+\frac{s^2}{2 m_H^4}\right)
\ea

In order to have the usual chiral power counting beginning at $\Opd$ for $N$
and $D$, we multiply both functions at the same time by $K$. This leaves
unchanged their ratio, $T(s)$, and their cut structure. Thus, we will have:

\ba
\label{otrand}
N(s)&=&K(s)\left(1+\delta N(s)\right) \\ \nonumber
D(s)&=&1+g(s)\, N(s)
\ea 

It is then easy to see that our $N$ function satisfies eq. (\ref{eqs2}) up to
one loop calculated at $\Opc$. In fact, at this level, $\Ima T_{Left}$ is given
by the imaginary part of eq. (\ref{kdeltaN}). The $D$ function satisfies eq.
(\ref{eqs1}) identically. This expresses that our final
amplitude $T(s)$, given in eq. (\ref{Tf}), is unitary to all orders.

It is interesting to compare eq. (\ref{Tf}) with the Inverse Amplitude
Method (IAM) \cite{IAM} which we have also used with great phenomenological 
success in the meson-meson scattering \cite{paquito}. The IAM has been as
well used in the
strongly interacting Higgs sector in ref. \cite{D1,D2,ramonet}. This method can be
obtained easily from eq. (\ref{Tf}) as a special case. To see this let us write:

\be
\label{k4}
N(s)=K(s)(1+\delta N(s))=N_2+N_4+...
\ee
with $N_2$ and $N_4$ the $\Opd$ and $\Opc$ contributions of the left hand side
respectively. Then expanding
$\left(K(1+\delta N)\right)^{-1}$ we have

\be
\label{kexp}
\frac{1}{K(1+\delta N)}=\frac{1}{N_2+N_4}=\frac{1}{N_2}-\frac{N_4}{N_2^2}+...
\ee

Introducing this result in eq. (\ref{Tf}), $T(s)$ reads

\be
\label{iam}
T(s)=\frac{1}{\frac{1}{N_2}-\frac{N_4}{N_2^2}+g(s)}=\frac{N_2^2}{N_2-N_4+N_2^2
g(s)}=\frac{T_2^2}{T_2-T_4}
\ee
with $T_2$ the $\Opd$ chiral amplitude and $T_4$ the $\Opc$ contribution, which
is given by $N_4-N_2^2 g(s)$. The last expression in eq. (\ref{iam})
is the usual way in which the IAM approach for a partial wave amplitude is 
presented. Thus, the IAM results in our formalism as a special case through 
the approximation given in eq. (\ref{kexp}).

Let us consider first the case $m_H<< 4\pi v\simeq 3$ TeV. In order to 
see analytically what is going on, let us note that for $s\lesssim m_H^2$ the 
local terms in eq. (\ref{kdeltaN}) of $\Opc$ divided by 
$(4 \pi v)^2$ are much smaller than the ones with $m_H^2$ in the denominator. 
On the other hand, close to the
bare pole, $s\simeq m_H^2$, the direct exchange in the s-channel of the resonance
dominates over their crossed exchanges and hence we will have:

\ba
\label{kap}
K(1+\delta N)&\approx& \frac{3 m_H^4}{2 v^2 (m_H^2-s)}\equiv
\frac{\alpha^2}{m_H^2-s} \\ \nonumber
T(s)&\approx& \frac{\alpha^2/(2 m_H)}{m_H-\sqrt{s}-\frac{i}{32 \pi m_H} \alpha^2}
\ea

If one considers $\alpha$ as an arbitrary parameter, only for 
$\alpha^2=3 m_H^4/(2 v^2)$ one has the SM.

Eq. (\ref{kap}) corresponds to a Breit-Wigner resonance, coming from the tree
level Higgs pole, with a mass $m_H$ and a width

\be
\label{W1}
\Gamma=\frac{\alpha^2}{16 \pi m_H}
\ee
For the SM Higgs boson one obtains the lowest order prediction

\be
\label{wmsm}
\Gamma=\frac{3m_H^3}{32 \pi v^2}
\ee
which is much smaller than $m_H$ for

\be
\label{smaller}
m_H<< 4\pi v \sqrt{\frac{2}{3 \pi}}\simeq 1.5 \hbox{TeV}
\ee

In fact, eq. (\ref{kap}) has sense only when $m_H<<\Gamma$. 

In the former situation, if we applied the IAM method we would obtain for
$s\approx m_H^2$

\be
\label{iam2}
T(s)=\frac{6 m_H^4/(11 v^2)}{\frac{6 m_H^2}{11} -s-i\frac{3 m_H^4}{88 \pi v^2}}
\ee
with corresponds to a Breit-Wigner with a pole at $\sqrt{s}=\sqrt{\frac{6}{11}}
m_H$ instead of $m_H$ as in eq. (\ref{kap}). Thus, the resummation done in eq.
(\ref{kexp}) for $\left[K(1+\delta N)\right]$ has shifted the tree level pole
from $m_H\rightarrow \sqrt{\frac{6}{11}} m_H \approx 0.74 m_H$. This possible 
source of inaccuracy of the IAM was already established in \cite{nd}. In the 
meson-meson scattering this does not occur because of
Vector Meson Dominance in the vector channels and the leading role of unitarity
together with coupled channels that happens in the resonant scalar channels with
I=0, 1 and 1/2 \cite{nd}. 

A much more interesting case occurs for $m_H>> 4 \pi v$. In this case, for
$\sqrt{s}\lesssim 4\pi v$, one can only retain in eq. (\ref{kdeltaN}) the two
first terms so that:

\be
\label{kap2}
K(s)(1+\delta N(s))\approx \frac{s}{v^2}-\frac{s^2}{16 \pi^2 v^4}\left(
a+b+\frac{7}{18}\log \frac{s}{m_H^2}\right)
\ee

In the former equation the $\Opd$ term is independent of the underlying
fundamental theory and is fixed by symmetry and the experimental value of
$v$. The same happens for the coefficient $7/18$ in front of 
${\displaystyle{\log\frac{s}{m_H^2}}}$ since it is given by loops at $\Opc$ in 
the crossed channels with the $\Opd$ amplitude at the vertices. However, the 
coefficients $a$ and 
$b$ depend on the underlying theory although one expects them to be of
$\mathcal{O}(1)$ since they are given over the relevant scale coming from loops
$4 \pi v$ \cite{Manohar}. In the case of the SM one has from eq. 
(\ref{kdeltaN}):

\be
\label{b}
b=\frac{1673-297 \sqrt{3} \pi}{108} \simeq \frac{1}{2}
\ee

However, the coefficient $a$ in eqs. (\ref{kdeltaN}) and (\ref{kap2}) is not 
fixed by the $\Opc$ chiral perturbation result since at this order cancels 
with the contributions proportional to $a$ from the $g(s)$ function, 
eq. (\ref{gs}). Substituting eq. (\ref{kap2}) in eq. (\ref{Tf}), $T(s)$ then 
reads

\be
\label{Tf3}
T(s)\simeq \left[\left(\frac{s}{v^2}-\frac{s^2}{16 \pi^2 v^4}\left(a+b+
\frac{7}{18}\log \frac{s}{m_H^2}\right) \right)^{-1}+\frac{1}{16 \pi^2}\left(
\log \frac{-s}{m_H^2}-a\right) \right]^{-1}
\ee

On the other hand, for $a$ and $b$ of $\mathcal{O}(1)$ and with $s\lesssim 1$
TeV$^2$, well below $(4 \pi v)^2$, the approximation given in eq. (\ref{kexp}) 
is numerically
accurate and then we recover the IAM result eq. (\ref{iam}). As a consequence 
our results from eqs. (\ref{Tf}) and eq. (\ref{kdeltaN}) coincide in this case 
with those already 
derived in \cite{D2} making use of the IAM. The most interesting conclusion is
that for $m_H \rightarrow \infty$ there is a pole in the I=L=0 partial wave
amplitude, with a vanishing mass and width as $m_H\rightarrow \infty$. Note also
that in eq. (\ref{iam}) the dependence on $a$ disappears. 

In order to understand properly the previous result one should keep in mind
that, in the present case, $m_H$ is just a scale \cite{D2} 
below which the only degrees of freedom are the Goldstone bosons. That is, there 
are no bare
resonances (elementary heavy fields) with any quantum numbers and masses 
below $m_H$. Taking this into
account, we can compare our result with the ones of ref. \cite{ramonet}. In 
this reference, the IAM is applied to study the resonance spectrum of the 
strongly interacting Higgs sector as a function of $L_1^r(\mu)$ and 
$L_2^r(\mu)$ with $\mu=1$ TeV. The case we are considering, $m_H>>4 \pi v$, 
corresponds to those values of the $L_i^r$ such that the underlying theory has not
bare resonances below $m_H$. For instance, from eq. (\ref{counter}) we see 
that, for $m_H>>4 \pi v$ and $\mu=1$ TeV, $L_1^r$ and $L_2^r$ are positive 
and large ($L_2^r\approx 2 L_1^r$)\footnote{For $m_H=150$ TeV, as considered in
Fig. 1, one obtains from eq. (\ref{counter}) $L_1^r=0.0024$ and $L_2^r=0.0043$,
with $\mu=1$ TeV. As one can see in Fig. 4 of ref. \cite{ramonet}, there is a 
rather narrow pole with a mass bellow 1 TeV in the I=L=0 channel for these 
values of the counterterms. This result is of course in agreement with the 
content of Fig. 1 of the present work.}. In fact in ref. \cite{ramonet}, for 
this region of values of the counterterms, one finds low
mass and narrow poles with I=L=0 and no resonances in the other channels 
(I=L=1 and I=2, L=0).

\begin{figure}[ht]
\centerline{
\protect
\hbox{
\psfig{file=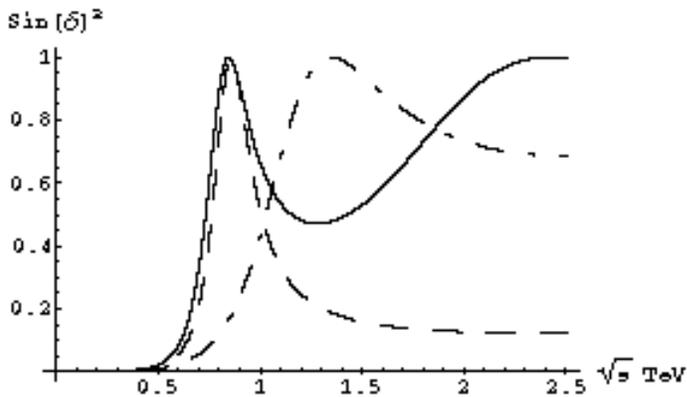,width=0.6\textwidth}}}
\caption{$|T(s)/(16\pi)|^2=(\sin\delta(s))^2$ where $\delta(s)$ is the phase 
of $T(s)$, eq.
(\ref{Tf3}). The solid line corresponds to $m_H=0.9$ TeV, the dashed 
line to
$m_H=150$ TeV and the dotted-dashed line to $m_H=11$ TeV. \it{Note as the
continuous curve ($m_H=0.9$ TeV) is very similar to that with $m_H=150$ TeV
(dashed line).}}
\end{figure}

In Fig. 1 we show three curves corresponding to $|T(s)/(16
\pi)|^2=(\sin\delta(s))^2$ with $\delta(s)$ the phase of $T(s)$, eq. 
(\ref{Tf}). Three 
different values for the parameter $m_H$ with $b$ given in eq. (\ref{b}) and 
$a=1$ are considered. Theses curves are
rather stable under changes of $a$ and $b$ of $\mathcal{O}(1)$, for instance, 
in the interval $[-1,1]$. The solid line is for $m_H=0.9$ TeV and one sees 
for
$s\lesssim 1.5$ TeV the presence of a clear resonance located around $m_H$, as
already discussed above in the case of $m_H<< 4\pi v$.
However, the dashed line, with $m_H=150$ TeV, also shows a clear signal 
for a 
narrow resonance at 1 TeV and with a shape around the pole position rather 
similar to that for $m_H=0.9$ TeV. This pole with a mass below 1.5 TeV 
begins to appear for 
$m_H\gtrsim 10$ TeV as already stated in \cite{D2}. The dotted-dashed line in 
fact corresponds to $m_H=11$ TeV and one sees clearly a bump
corresponding to this pole.
However, one has to realize that this state appears {\bf{without}} the tree 
level pole of the Higgs boson, as it is expected from eq. (\ref{Tf3}) and as 
we have explicitly checked from eqs. (\ref{Tf}) 
and (\ref{kdeltaN}) by removing in the last equation the last line. As a 
consequence this state does not originate from the tree level Higgs pole and 
is not a Higgs boson responsible for the spontaneous electroweak symmetry
breaking. It 
is just a consequence of the strong interactions between the $W_L$ bosons 
giving rise to a $W_L W_L$ resonance, similarly as
the deuteron pole is a bound state of a proton and a neutron. 

 For $m_H\approx 4 \pi v$ the physics involved is much more dependent on the
 values of the parameters $a$ and $b$ and both the tree level pole of the Higgs
 and the rest of counterterms are similarly important and have large 
 interferences. In
 general, as suggested by eq. (\ref{wmsm}), the Higgs is very wide, with a 
 width as large or more than its mass. This is in fact what happens for 
 instance for 
 $b$ given by
 eq. (\ref{b}), $m_H=3$ TeV and $a=0$, where a pole coming from the Higgs tree 
 level pole 
 appears around $(1.2 + i\, 0.6)$ TeV in the unphysical sheet. However, 
 for certain values of $a$ of $\mathcal{O}(1)$, for instance, $a=-2$ one sees 
 two poles at: $(4+i\, 0.04)$ and  $(1.2+i\, 0.7)$ TeV. As a consequence, the 
 deviations
 from the light Higgs situation can be very large and simple Breit-Wigner 
pictures from bare poles are not adequate in this case since the naive formula
 given in eq. (\ref{wmsm}) gives a width that for $m_H>1.5$ TeV is larger than
$m_H$. This is a clear signal that the situation cannot be reduced to such 
simple terms which are valid only for narrow resonances ($\Gamma<<m_H$).

In ref. \cite{Igi} the N/D method is also used to study the strongly
interacting Higgs sector. We reproduce their results with a suitable choice of
the constant $a$. However, we would like to indicate that while we 
maintain here the chiral power counting and recover the full chiral amplitude 
up to $\Opc$ this is not the case in \cite{Igi}. Another difference is the way
in which the left hand cut is treated. We have included it in a chiral loop 
expansion while in that reference the left hand cut is fixed {\it{a priori}} 
to be given by the crossed Higgs exchange.

\vspace{1cm}
{\bf{Conclusions}}

Making use of the N/D method and the electroweak effective chiral Lagrangians we
have studied the properties of the strong $WW$ scattering in the scalar channel.
This approach gives rise to physical full unitarized amplitudes respecting the
low energy symmetry constraints. In particular, it is able to describe
resonances as poles in the unphysical sheet. In this way, we have also played 
a special attention to the nature of the resonances relevant within the LHC 
energy range. There are two clearly
distinct cases: For $m_H<< 4\pi v$ the amplitudes are dominated by the tree
level Higgs boson pole. For $m_H>> 4\pi v$, although the previous tree level 
Higgs pole disappears for the energies considered, there is another
physical pole below $\sqrt{s}<1.5$ TeV. The shape of this resonance can be 
very similar to that of the first case. However, the nature of this second pole 
is completely different, since it corresponds to a dynamical $WW$ resonance 
and hence it is not responsible for the spontaneous breaking of the symmetry. We
have also seen that this conclusion is stable under changes of the chiral
parameters of higher order. Thus, it should be a common
feature of any underlying theory responsible for the electroweak symmetry
breaking with unbroken custodial symmetry and with heavier particles appearing
with a mass $m_H>> 4\pi v$.

\vspace{0.4cm}

{\bf Acknowledgments}

I would like acknowledge a critical reading and fruitful discussions with J. R.
Pel\'aez, E.
Oset and M. J. Vicente-Vacas. This work has been supported by an FPI 
scholarship of the Generalitat Valenciana and partially by the EEC-TMR 
Program$-$Contact No. ERBFMRX-CT98-0169.

\end{document}